# Freezing of Gait as a Complication of Pallidal Deep Brain Stimulation in DYT-*KMT2B* Patients with Evidence of Striatonigral Degeneration

*Laura Cif, MD, PhD,*[1,*] *Diane Demailly, MD,*[1] *Xavier Vasques, PhD,*[2,3] *Delphine de Verbizier, MD,*[4] *Philippe Coubes, MD, PhD,*[1] *Kathleen Gorman, MD,*[5,6] *and Manju A. Kurian, PhD*[5,6]

Mutations in *KMT2B* were first identified in individuals with early-onset complex dystonia.[1] Since then, it is emerging as one of the most common causes of genetic childhood-onset dystonia.[2] Additional features include short stature, dysmorphism, developmental delay, psychiatric features, endocrinopathy and others.[3] Patients with DYT-*KMT2B* refractory generalized dystonia maintain significant benefit from internal globus pallidus deep brain stimulation (GPi-DBS) therapy as previously reported, except for laryngeal dystonia and gait.[3] We wish to contribute our observation of freezing of gait (FOG) in *KMT2B*-related dystonia by reporting five subjects (four females) with protein truncating variants (PTV) (Table 1).

Clinical presentation and response to DBS was previously reported.[3] The mean age at dystonia onset was 3.6 years (range: 2–6 years). The median age at DBS implant was 23 years (IQR: 8–30.3 years), and patients were followed for a median of 14.5 years (IQR: 8.5–24 years) after DBS. As illustrated in the Videos 1, 2, and 3, FOG was documented in all, occurring from 14–43 years of age (range: 1–15.5 years after GPi-DBS). DaTscan (SPECT for $^{123}$I-ioflupane) was abnormal in 4/5 patients undertaken from 2.5 years before, to 24 years after DBS insertion (Table 1). Subject 3, despite normal DaTscan at age 20.5 years (2.5 years pre-DBS), exhibited mild FOG when turning with lower limb dystonia from 1 year after DBS insertion. DaTscan repeated 13 years later in subject 3 identified bilateral decrease of putaminal uptake. Prior to DBS, dystonia was unresponsive to L-dopa in all subjects, as was FOG post-DBS in 2/5. Only 1/5 has maintained independent gait at last follow-up, despite 4/5 having recovered autonomous gait at steady state under DBS.

Feuerstein et al.[4] reported on the emergence of parkinsonism and abnormal brain DaTscan imaging in a patient with a heterozygous loss-of-function *KMT2B* variant (c.974_979del, p.Ser325⋆). He presented at aged 3 years with typical features of DYT-*KMT2B*, necessitating GPi-DBS aged 23 years. Though, initial benefit was evident, by 33 years, generalized dystonia, parkinsonism with rigidity, bradykinesia and FOG predominated. Extensive DBS reprogramming and switch-off did not modify symptoms. DaTscan showed bilateral decreased putaminal uptake. In this patient, Rotigotine (but not L-dopa) significantly improved FOG.

In our DYT-*KMT2B* group, FOG occurred across a broad age range, from an early post-operative presentation to more than 15 years after DBS insertion. All patients had PTVs. In a previous publication, in DYT-*KMT2B*, dystonia severity scores appeared to be comparable and more severe in PTVs and chromosomal deletions versus missense variants, possibly suggestive of a relationship between motor severity and type of *KMT2B* variant.[3] To date, FOG has not been reported in patients with missense variants; identification of further cases will determine whether this is a true genotype–phenotype correlation.

Many dystonia-parkinsonism disorders are associated with reduced striatal dopamine due to degeneration of substantia nigra

[1]Département de Neurochirurgie, Unité des Pathologies Cérébrales Résistantes, Unité de Recherche sur les Comportements et Mouvements Anormaux, Hôpital Gui de Chauliac, Centre Hospitalier Universitaire Montpellier, Montpellier, France; [2]Laboratoire de Recherche en Neurosciences Cliniques, Montpellier, France; [3]IBM Technology, Bois-Colombes, France; [4]Département de Médecine Nucléaire, Hôpital Gui de Chauliac, Centre Hospitalier Universitaire Montpellier, Montpellier, France; [5]Developmental Neurosciences, UCL Great Ormond Street Institute of Child Health, Zayed Centre for Research into Rare Disease in Children, London, United Kingdom; [6]Department of Neurology, Great Ormond Street Hospital, London, United Kingdom

*Correspondence to: Dr. Laura Cif, Département de Neurochirurgie, Hôpital Gui de Chauliac, Centre Hospitalier Universitaire Montpellier; Unité de Recherche sur les Comportements et Mouvements Anormaux; 80, Avenue Augustin Fliche; 34000 Montpellier, France; E-mail: a-cif@chu-montpellier.fr

Keywords: *KMT2B*, freezing of gait, DaTscan, deep brain stimulation.

TABLE 1 *Genotype, demographics, age at dystonia onset, at freezing of gait (FOG) and at DBS insertion and follow-up with DBS*

| Subject | Variant inheritance | Current age sex | Dystonia onset (yr) | Dystonia at onset → evolution | Age at GPi DBS (yr) FU with DBS (yr) | Age FOG (yr) post-DBS FOG (yr) | Age at DaTscan (yr) | DaTscan index/ descriptive | DaTscan index/ descriptive | DaTscan/ MRI |
|---|---|---|---|---|---|---|---|---|---|---|
| 1 9 | c.1656dupC p.Lys553Glnfs*46 de novo | 29.5 F | 3 | LL dystonia → generalized dystonia with cranial, cervical & laryngeal | 5 23 | 20.5 15.5 | 17 | *Right:* St: 2.46 Pu: 2.47 Cau: 3.02 Bilateral short putamen | *Left:* St: 2.23 Pu: 2.15 Cau: 2.61 | 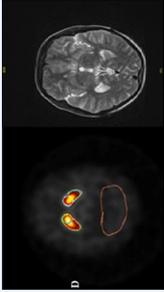 |
| 2 10 | c.2137dupA p.Thr713Asnfs*4 Absent in mother | 33 F | 3 | LL dystonia → generalized dystonia with cranial, cervical & laryngeal | 8 24 | 14 6 | 33 | *Right:* St: 1.8 Pu: 1.2 Cau: 2.5 Bilateral putaminal hypofixation | *Left:* St: 2 Pu: 1.3 Cau: 2.8 | 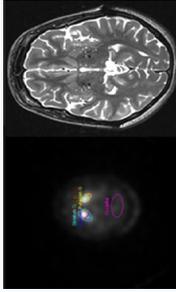 |
| 3 17 | c.3147_3160 del p.Gly1050Profs*33 Inherited from symptomatic mother | 33.5 M | 4 | Laryngeal dystonia → LL, laryngeal & cervical | 23 10 | 24 1 | 20.5 | *Right:* St: 2.87 Pu: 2.75 Cau: 3.08 Normal | *Left:* St: 3.05 Pu: 3.19 Cau: 3.04 | 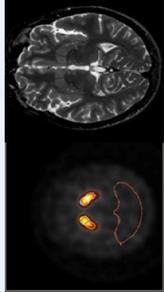 |

(*Continues*)

**TABLE 1** Continued

| Subject[^] | Variant inheritance | Current age | sex | Dystonia onset (yr) | Dystonia at onset → evolution | Age at GPi DBS (yr) FU with DBS (yr) | Age FOG (yr) post-DBS FOG (yr) | Age at DaTscan (yr) | DaTscan index/descriptive | DaTscan/MRI |
|---|---|---|---|---|---|---|---|---|---|---|
| 4 21 | c.3642 + 5G > A de novo | 45 | F | 6 | LL dystonia → generalized dystonia with cranial, cervical & laryngeal | 28 17.5 | NA NA | 45 | *Right:* St: 1.9 Pu: 1.6 Cau: 2.3 Bilateral short putamen | *Left:* St: 2 Pu: 2.1 Cau: 2.2 | 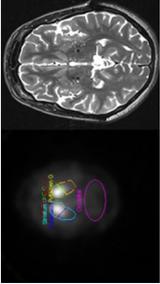 |
| 5 37 | c.6439C > T p.Gln2147* de novo | 45 | F | 2 | LL dystonia → generalized dystonia with cranial, cervical & laryngeal | 37 8.5 | 43 6 | 36 | *Right:* St: 1.87 Pu: 1.87 Cau: 2.29 Bilateral short putamen | *Left:* St: 1.87 Pu: 1.91 Cau: 2.18 | 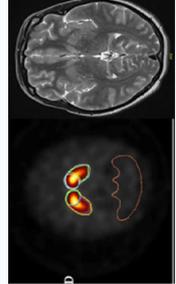 |

SPECT for $^{123}$Ioflupane (DaTscan) imaging and MRI.
^Refers to number in Cif et al.[3]
Abbreviations: Cau, Caudate; F, female; LL, lower limb; M, male; Pu, putamen; St, striatum. NA—not able to walk since severe feet deformities.

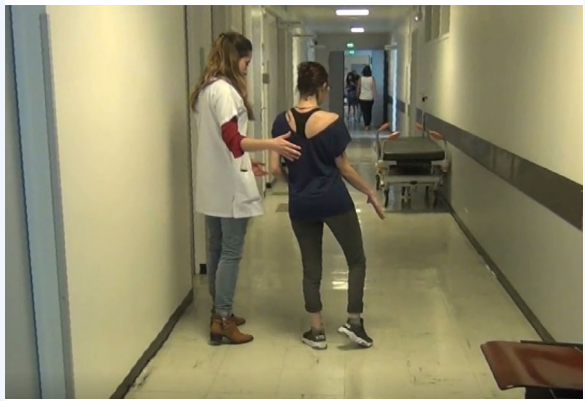

**Video 1.** Freezing of gait (FOG) in DYT-*KMT2B* under GPi-DBS. Subject 1 in the manuscript: The two video sequences (S1, S2) are post-DBS. Sequence 1 documents FOG as a persistent feature despite adjustment of DBS settings at age 20.5 and sequence 2 age 29.5 years respectively, 15 and 24 years after DBS initiation. Subject 2 in the manuscript: The two video sequences (S3, S4) are recorded post operatively, under DBS. Sequence 3 documents FOG at age 14 and sequence 4 at age 31 respectively, 6 and 17 years after DBS initiation.
Video content can be viewed at https://onlinelibrary.wiley.com/doi/10.1002/mdc3.13519

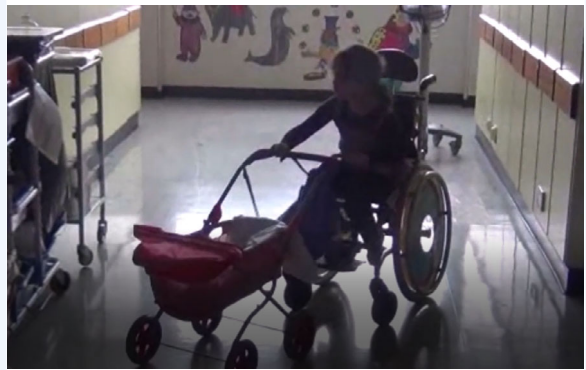

**Video 3.** Additional subjects (not included in the manuscript): Subject 1: four consecutive video sequences document gait evolution after DBS; gait was not available pre-DBS since she was in *Status Dystonicus*; just after DBS insertion, she was still wheelchair bound, unable to stand and to walk; lower limb dystonia improved allowing standing and gait with support, without FOG. Subject 2: two consecutive video sequences are presented; the first sequence documents dystonic features involving lower limbs during gait pre-DBS; the second sequence shows occurrence of FOG early after DBS insertion.
Video content can be viewed at https://onlinelibrary.wiley.com/doi/10.1002/mdc3.13519

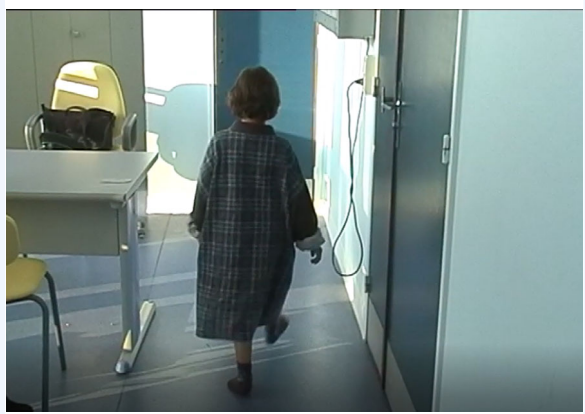

**Video 2.** Video sequences for the patients included in the manuscript illustrating gait previous to the occurrence of FOG. Sequence 1: illustrates subject 1 from the manuscript, early after DBS insertion; lower limbs dystonia is improved compared to preoperative assessment, allowing standing and walking; gait is still impaired by residual dystonia, but the patient does not present yet FOG; the two first video sequences from Video 1 illustrate the occurrence and worsening of FOG over time under DBS for the patient. Sequences 2 and 3: illustrate subject 4 from the manuscript, pre-DBS and with DBS, respectively. Despite obvious significant improvement of her dystonia under DBS, she was unable to walk because of severe skeletal deformities involving the lower limbs. Sequences 4, 5 and 6: illustrate subject 5 from the manuscript, pre-DBS and after DBS insertion, before the occurrence of FOG. In sequence 4 (pre-DBS), gait is very unsteady because of severe dystonia involving trunk and lower limbs; sequence 5 documents significant improvement under DBS, without FOG. Sequence 6 shows altered cadence with mild FOG at turn around.
Video content can be viewed at https://onlinelibrary.wiley.com/doi/10.1002/mdc3.13519

pars compacta neurons.[5] Our finding of bilateral short putamen on DaTscan is suggestive of striatal dopaminergic denervation in DYT-*KMT2B*. Contrary to levodopa-induced dyskinesia and wearing-off phenomena in Parkinson's disease (PD), the pattern of striatal dopamine depletion does not seem to affect the risk of FOG in PD.[6] Nevertheless, in de novo PD patients, those with severe reduction of DaT uptake in the caudate and putamen have a significantly higher incidence of FOG than those with mild or moderate uptake reduction.[7] Therefore, it is possible that the specific striatal anatomy and reduced dorsolateral (motor) putaminal DaT uptake in DYT-*KMT2B* patients could potentially propagate network alterations to drive the DBS-mediated switch to hypokinesia, despite ongoing therapeutic effect for dystonia.

The relationship between DYT-*KMT2B*, FOG and DBS intervention remains yet to be fully elucidated; nevertheless, and contrary to other monogenic dystonias, in DYT-*KMT2B*, DaTscan abnormalities and FOG is an emerging phenomenon, at least in those patients with PTVs. Long-term GPi-DBS is reported to lead to hypokinetic gait disorders in patients treated for dystonia.[8] In DYT-*TOR1*A, physiological parameters such as the paired associative stimulation response were almost absent and short-interval intracortical inhibition reduced.[9] This pattern, resembles untreated PD may in part explain the observed hypokinesia in DBS-treated dystonia without abnormal DaTscan imaging.[8]

In conclusion, the potential risk of hypokinetic gait disorders in DYT-*KMT2B* should be considered in patients undergoing GPi-DBS, which warrants strict monitoring of the motor phenomenology post-procedure. Serial DaT SPECT imaging may aid identification of striatal dopaminergic denervation in DYT-*KMT2B* and a clinical trial of levodopa or dopaminergic agonist may be useful. More evidence is needed to improve understanding of the etiological basis and efficacy of different interventions for DYT-*KMT2B*-related hypokinetic movement disorders.

## Author Roles

(1) Research project: A. Conception, B. Organization, C. Execution; (2) Statistical Analysis: A. Design, B. Execution, C. Review and Critique; (3) Manuscript: A. Writing of the first draft, B. Review and Critique.

LC: 1A, 1B, 1C, 2A, 2C, 3A, 3B.
DD: 1C, 2C, 3B.
XV: 1C, 2B, 2C, 3B.
DDV: 1C, 2C, 3C.
PC: 1C, 2C, 3C.
KG: 1A, 1C, 2B, 2C, 3A, 3B.
MAK: 1A, 1C, 2B, 2C, 3A, 3B.

## Disclosures

**Ethical Compliance Statement:** The study was approved by the Internal Review Board of Montpellier University Hospital (Ethics Board number 2018_IRB-MTP_11–11). Written informed consent was obtained for all participants in whom research genetic testing was undertaken and for publication of videos. We confirm that we have read the Journal's position on issues involved in ethical publication and affirm that this work is consistent with those guidelines.

**Funding Sources and Conflicts of Interest:** No specific funding was received for this work. The authors declare that there are no conflicts of interest relevant to this work.

**Financial Disclosures for the Previous 12 Months:** L.C., D.D., X.V., D.D.V., P.C., K.G. and M.A.K. declare that there are no additional disclosures to report. ■